# Anisotropic defect-induced ferromagnetism and transport in Gd-doped GaN two-dimensional electron gasses

Zihao Yang,[1] Thomas F. Kent,[2] Jing Yang,[2] Hyungyu Jin,[3] Joseph P. Heremans,[3,4] and Roberto C. Myers[1,2,4]

[1]Department of Electrical and Computer Engineering, The Ohio State University, Columbus, Ohio 43210, USA

[2]Department of Materials Science and Engineering, The Ohio State University, Columbus, Ohio 43210, USA

[3]Department of Mechanical and Aerospace Engineering, The Ohio State University, Columbus, Ohio 43210, USA

[4]Department of Physics, The Ohio State University, Columbus, Ohio 43210, USA

## Abstract

Here we report on the effect of rare earth Gd-doping on the magnetic properties and magnetotransport of GaN two-dimensional electron gasses (2DEGs). Samples are grown by plasma-assisted molecular beam epitaxy and consist of AlN/GaN heterostructures where Gd is delta-doped within a polarization-induced 2DEG. Ferromagnetism is observed in these Gd-doped 2DEGs with a Curie temperature above room temperature and an anisotropic spontaneous magnetization preferring an out-of-plane (c-axis) orientation. At magnetic fields up to 50 kOe, the magnetization remains smaller for in-plane configuration than for out-of-plane, which is indicative of exchange coupled spins locked along the polar c-axis. The sample with the lowest Gd concentration (2.3 × $10^{14}$ cm$^{-2}$) exhibits a saturation magnetization of 41.1 $\mu_B$/Gd$^{3+}$ at 5 K revealing that the Gd ion spins (7 $\mu_B$) alone do not account for the magnetization. Surprisingly, control samples grown without any Gd display inconsistent magnetic properties; in some control samples weak ferromagnetism is observed and in others paramagnetism. The ferromagnetic 2DEGs do not exhibit the anomalous Hall effect; the Hall resistance varies non-linearly with the magnetic field, but does not track the magnetization indicating the lack of coupling between the ferromagnetic phase and the conduction band electrons within the 2DEG.

# I. Introduction

The III-nitride family semiconductors are good host materials for the magnetic rare earth element gadolinium (Gd) because of its high solubility (dilute doping regime) and its ability to form rocksalt GdN in wurtzite GaN (heavy doping regime).[1,2] Further, the $Gd^{3+}$ ion contains 7 unpaired f-shell electron spins that open the possibility of engineering magnetism into nitride semiconductors. In particular, in the dilute doping regime, an anomalous ferromagnetic phase with a Curie temperature ($T_C$) above room temperature has been reported by multiple groups, with the magnetism being thought to arise from the crystallographic point or line defects.[1,3-5] This room temperature ferromagnetic phase may point to a new magnetic semiconductor material system that overcomes the disadvantage of $T_C$ being well below room temperature in the prototypical magnetic semiconductor GaMnAs, which limits its practical applications.[6] However, to the best of our knowledge there are no current reports on the electrical transport properties in conjunction with the magnetic properties of the Gd-doped GaN or its related structures. Lo et al. reported on the anomalous Hall effect in Gd ion-implanted GaN/AlGaN 2DEGs, however the magnetic properties were not measured.[7] Therefore the possible utility of Gd:GaN as a useful ferromagnetic semiconductor remains dubious.

In this work, we examine the interaction between the anomalous ferromagnetic phase in Gd-doped GaN and the conduction band electrons via magnetotransport measurements. By delta-doping Gd directly into a polarization-induced AlN/GaN two-dimensional electron gas (2DEG), a high density of electrons is spatially overlapping with the magnetically doped region. Delta-doping also allows for exploration of much higher Gd-doping densities than previously explored. Here we dope the 2DEG region with up to $4.6 \times 10^{14}$ cm$^{-2}$ Gd. In these 2DEGs, delta-doping with Gd is observed to induce a ferromagnetic phase with $T_C$ above room temperature and the saturation magnetization is much larger than can be accounted for by Gd ions spins alone. An out-of-plane (c-axis) easy-axis is observed in Gd-doped samples which is in contrast to the previous report by Pérez et al.[8] More interestingly, at magnetic fields up to 50 kOe, the magnetization remains smaller for in-plane configuration than for out-of-plane, which is indicative of exchange coupled spins locked along the polar c-axis. Magnetotransport measurements demonstrate the presence of the 2DEG in Gd-doped and un-doped samples where the carrier concentration of these samples are found to be on the order of 10$^{13}$ cm$^{-2}$. Given that the Gd-doping is spatially confined to the 2DEG, one might have expected that magnetic scattering or exchange coupling of the 2DEG and the magnetic phase would lead to an anomalous Hall effect (AHE). However, the measured non-linear component $\rho_{xy}^{NL}$ in the Hall resistivity does not display ferromagnetic behavior, but instead results from the effect of the magnetoconductivity $\sigma_{xx}$, itself due to the effect of weak localization, on Hall resistivity $\rho_{xy}$. We observe a decrease in the electron mobility as the Gd-doping concentration is increased suggesting that whatever defects forms as a result of Gd-doping, they also lead to strong Coulombic charge scattering on Gd generated defects. The lack of the ferromagnetic response in the Hall resistivity is consistent with a defect-induced ferromagnetic phase that is uncoupled to the conduction band of GaN. This lack of coupling is perplexing given that the magnetic doping is performed precisely within the 2DEG. We speculate that the ferromagnetic phase consists of high densities of

Gd-associated defects (lines or clusters) that locally deplete the 2DEG, thereby preventing any significant coupling between the 2DEG and the ferromagnetic spins. The observation of lack of coupling also agrees well with a recent study by Buß et al. where no exchange coupling between the conduction band electrons and the Gd ions or Gd associated polarized lattice ions is found in Gd ion-implanted GaN thin films using time-resolved magneto-optical spectroscopy.[9] Our results indicate that Gd-doped GaN cannot be considered as a ferromagnetic semiconductor as far as its magnetotransport properties are concerned.

## II. Experiment

### A. Sample growth

All samples are grown by plasma-assisted molecular beam epitaxy (PAMBE) on a Veeco GEN930 PAMBE system with a base pressure of $5 \times 10^{-11}$ torr.[2,10] The PAMBE system is equipped with effusion cells of Al (6N5 purity), Ga (7N purity), Gd (4N7 purity) and a nitrogen plasma source (6N purity). To begin, the Gd-doping rate is calibrated by growing a stack of Gd:GaN with various Gd effusion cell temperatures. The initial Gd:GaN layer is grown on a GaN buffer layer that was grown on an AlN on sapphire (Kyma) template with a Gd effusion cell temperature of 900 $°C$ for 30 minutes. The subsequent Gd:GaN layers are grown with Gd effusion cell temperatures ranging from 950 $°C$ to 1150 $°C$ with 50 $°C$ steps for 15 minutes. Ten-minute wait times are used between Gd cell temperature changes. The Gd concentration is measured using a PHI6600 Quadrupole secondary ion mass spectrometry (SIMS) instrument with 5 keV $O_2$ primary ion bombardment, and calibrated by Rutherford backscattering spectroscopy (RBS) on a Gd ion implanted standard sample with an implant dose of $1.01 \times 10^{15}$ cm$^{-2}$. The Gd-doping concentration profiles shown in Fig. 1a are calculated separately for each of the five major Gd isotopes (thin colored lines), and then averaged to provide a final Gd-doping profile (thick black line). The extracted average Gd-doping concentration (black dot) as a function of the reciprocal of the Gd cell temperature is shown in Fig. 1b and can be fitted with the Clapeyron equation $n_{Gd} \propto p_{Gd} = A \times e^{-b/T}$ where $n_{Gd}$ is the Gd-doping concentration and $p_{Gd}$ is the equilibrium partial pressure of the Gd beam.[11] The Gd-doping rate (in units of effective Gd-delta-doping monolayer coverage (ML) per second) is then calculated and shown in Fig. 1c by assuming that GdN remains in the wurtzite structure for small doping levels.[2] All samples in this report utilize a growth rate of 0.007 ML/s corresponding to a Gd effusion cell temperature of 1169.4 $°C$.

Two sets of 2DEG samples, as shown in Fig. 2a, are grown on 6H-SiC (Si-face) wafers with different Gd-doping concentrations consisting of the following layers: 5 nm AlN (Al-face) capping layer/0.4 nm GaN spacer/0 ML or 0.2 ML ($N_{Gd}^{2D}$=0.2 ML) or 0.4 ML ($N_{Gd}^{2D}$=0.4 ML) Gd-delta-doping/300 nm smooth unintentionally doped (UID) GaN (Ga-face)/140 nm rough UID GaN/70 nm UID AlN nucleation layer/6H-SiC wafer. The samples without Gd-doping (0 ML) are denoted as Control Sample 1 (CS 1) or 2 (CS 2). We utilize identical growth conditions as developed in Kent et al.[2] for the Gd-delta doping, where it was shown that at a Gd dose of greater than 1.2 ML, rocksalt GdN nanoislands form. By remaining well below the critical dose in this study only single phase wurtzite Gd:GaN forms. To realize the structures, the growth chamber pressure is kept at $2 \times 10^{-5}$ torr by regulating the nitrogen flow rate. The nitrogen limited growth rate of GaN is 250 and 290 nm/hour for the first and second sets of samples respectively, and the nitrogen limited growth rate of AlN is 0.8% less than the GaN growth rate. The AlN nucleation layer is first grown at 780 $°C$ under nitrogen-rich regime with Al to N flux ratio of 0.6. The substrate temperature is then changed to 730 $°C$ to grow the rough GaN layer with Ga to N flux ratio of 1.7 under Ga-rich intermediate regime. A smooth GaN layer is subsequently grown under the droplet regime at 710 $°C$ with Ga to N flux ratio of 2.6. After the growth of the 300 nm smooth GaN layer, the Ga shutter is closed and the Gd shutter is opened for 30 seconds (0.2 ML Gd-doping) or 1 minute (0.4 ML Gd-doping). During this period, only Gd and N are being deposited. To ensure good heterointerface quality, a 0.4 nm GaN spacer is grown after the deposition of Gd. The final AlN capping layer is grown under a slightly Al-rich intermediate regime with Al to N flux ratio of around 1 at 710 $°C$.

### B. Band diagram and structural properties

The band diagram of the control samples is simulated using a 1-dimensonal Poisson-Schrodinger solver.[12] The AlN surface pinning level is set to 3.1 eV (mid-gap) below the conduction band minimum and the background doping in the smooth GaN layer is assumed to be donor-like with a concentration of $10^{16}$ cm$^{-3}$ due to the oxygen impurities.[13] The dotted band diagram and carrier concentration profile depicted in Fig. 2b is simulated under the assumptions that the AlN layer is fully strained and the conduction band offset ($\Delta E_c^0$) between the AlN and GaN is 2.1 eV.[13] The fully-strained AlN layer gives a strong piezoelectric charge density that induces a sharp band bending in the AlN layer resulting in a deep triangular quantum well with a 2DEG concentration of $4.2 \times 10^{13}$ cm$^{-2}$.

To demonstrate the spatial overlap between the 2DEG and Gd dopants, the quantitative Gd concentration profile is examined by SIMS using CsM+ primary ions with impact energy of 2 keV. The calibrated SIMS depth profile for Gd is shown in Fig. 2c showing that the concentration peaks at 0.4 nm away from the AlN and GaN interface overlapping with the 2DEG with a Gd concentration of $2.1 \times 10^{14}$ cm$^{-2}$, corresponding to 0.2 ML equivalent ($2.3 \times 10^{14}$ cm$^{-2}$) of Gd dopant atoms. Although the Gd doping profile extends into the smooth GaN region, the majority of Gd dopants are confined within the quantum well region. The Ga, Al, and N concentrations are raw, uncalibrated SIMS signal serving only as a marker for the 2DEG region. The apparent reduction of Al concentration near the surface is a SIMS measurement artifact.

High resolution X-ray diffraction (XRD) data on the Control Sample 1 are shown in Fig. 3a. The rocking curve full widths at half maximum (FWHM) for the GaN (002) and GaN (102) diffraction peaks are 792 arcsec and 1116 arcsec, respectively, which are higher than other reported values.[14,15] From the rocking curve FWHM, the density of screw-type threading dislocations can be estimated using the Hirsch model,[16-18]

$$D_S = \frac{\beta_{002}^2}{4.35 \times |b_s|^2} \quad (1)$$

where $|b_s|$ is the Burgers vector of the screw-type ($|b_s|$=0.5185 nm) dislocations, and $\beta_{002}$ is the FWHM of the (002) rocking curve. By comparison to the rocking curve data, the screw-type dislocations density is estimated to be $6.3 \times 10^8$ cm$^{-2}$. When the FWHM of the (002) and (102) are comparable, the screw and edge component of the threading dislocations have similar densities.[17] Since the ratio of (002) and (102) rocking curve FWHM in our study is 0.7, then the edge threading dislocation density should be similar to that of the screw-type dislocations, i.e. ~ $10^9$ cm$^{-2}$.

Representative atomic force microscopy (AFM) images from the 0.2 ML and 0.4 ML Gd-doped samples are shown in Fig. 3b and c. The surface morphologies consist of terrace-like features consistent with step-flow growth regime of the smooth GaN layer and AlN capping layers. The average surface roughness of both samples (0.56 nm and 0.43 nm) is comparable to that of the control samples (0.5 nm).

### III. Experimental Results

#### A. Magnetic Properties

The magnetic properties are analyzed by superconducting quantum interference device (SQUID) mangetometry using a Quantum Design MPMS XL. The room temperature magnetization data illustrated in Fig. 4a have had the dominant diamagnetic background subtracted and normalized by the sample surface area due to the nature of the 2D doping. The sample doped with 0.2 ML Gd possesses a room temperature ferromagnetic response characterized by a sharp increase in the magnetization at low magnetic fields (< 5 kOe), followed by full saturation up to a value of $6.5 \times 10^{-7}$ emu/mm$^2$ at 20 kOe. To examine the role of Gd-doping on this ferromagnetic phase, the results are compared with identical measurements on a 6H-SiC wafer, Control Samples 1 and 2. The out-of-plane magnetization data from the 6H-SiC wafer shows a very weak ferromagnetic response with a magnetization of less than $1 \times 10^{-7}$ emu/mm$^2$, indicating the dominance of the diamagnetic response (Fig.

4a top left inset). The weak marginal ferromagnetism in 6H-SiC is reportedly induced by defects in SiC.[19] The volume diamagnetic susceptibility of the SiC is calculated as $1.17 \times 10^{-5}$ cm$^3$/mol, very close to the reported value of $1.06 \times 10^{-5}$ cm$^3$/mol (c.g.s unit).[20] Surprisingly, our non-Gd-doped 2DEG, Control Sample 1, shows a stronger ferromagnetic behavior with a saturation magnetization of $1.9 \times 10^{-7}$ emu/mm$^2$ compared to SiC which could be explained by the theoretical modeling in Ref. 21 where the authors suggested that point defects (cation vacancies that act as acceptors) in wide bandgap semiconductors (such as GaN and BN) exhibit a moment of 3 $\mu_B$ per vacancy with long-rang coupling. The second non-Gd-doped 2DEG, Control Sample 2, exhibits a weak ferromagnetic response that is comparable to that of the bare SiC wafer, suggesting that the ferromagnetism in Control Sample 2 originates from the SiC substrate and not the GaN region (Fig. 4a top left inset). The data are assumed to be free of transition metal contamination since care was taken in sample preparation (cleanroom environment, plastic tweezers, HCl acid cleaning, and organic solvent cleaning). Combining the magnetization data at both 5 K (Fig. 4a bottom right inset) and 300 K for the control samples and the SiC wafer, we conclude that not only do Gd-doped GaN heterostructures exhibit ferromagnetism, but also that unintentionally doped GaN heterostructures contain free spins which can either be uncoupled and exhibit paramagnetic behavior at 5 K (Control Sample 2) or ferromagnetically coupled (Control Sample 1). The difference in the magnetic response between the control samples is attributable to variation in the density of these spin polarized defects. The magnetization of the 0.4 ML Gd-doped sample is only slightly greater than Control Sample 2, while the lighter doped sample exhibits a much stronger magnetization. Similar large variation in the magnetization of Gd-doped GaN was also observed by Roever et al.[22] In the current sample sets, we do however observe a boost in this defect magnetism upon Gd-doping, which follows previous reports on Gd-doped GaN.[1,23]

Magnetic hysteresis loops of the Gd-doped 2DEG samples are shown in Fig. 4b at two different temperatures measured with the field applied out-of-plane (parallel to the c-axis). With the magnetization normalized by total Gd content, the spontaneous magnetization at 20 kOe varies from around 30 $\mu_B$/Gd$^{3+}$ to 38 $\mu_B$/Gd$^{3+}$ at 300 K and 5 K, respectively for the sample with 0.2 ML Gd-delta-doping (50 kOe data is also shown in Tab. I). As these values are greater than the atomic moment of a single Gd atom (7 $\mu_B$/Gd$^{3+}$), then clearly the Gd ions are not the major contributor (if at all) to the ferromagnetic phase. This observation is not entirely surprising considering that a ferromagnetic phase occurs even in samples lacking any Gd-doping (Control Sample 1, see above). In Ref. 23, it was concluded that an acceptor-like defect band existed, based on hopping transport measured in Gd:GaN. At Gd-doping levels of $2 \times 10^{16}$ cm$^{-3}$, $6 \times 10^{18}$ cm$^{-3}$ acceptor-like defects were observed, indicating that Gd-doping is catalytically generating spin polarized defects. Other possible magnetic defects have been proposed including interstitial nitrogen, oxygen in octahedral sites, and Ga vacancies.[4,5] However, Roever et al. ruled out gallium vacancies and gallium vacancy clusters, and suggested that extended defects likely play a role in the ferromagnetism.[22] For the sample with 0.4 ML doping, the normalized magnetization at 20 kOe decreases to 7 $\mu_B$/Gd$^{3+}$ and 8.4 $\mu_B$/Gd$^{3+}$ at 300 K and 5 K, respectively (50 kOe data is also shown in Tab. I). This decrease in normalized magnetization with increasing Gd concentration is in agreement with previous reports on Gd-doped GaN, from which we conclude that the Gd contribution to the anomalous ferromagnetic phase is insignificant.[1,22] In both samples, neither the saturation moment nor the coercive field (with in-plane magnetic field) display significant changes from 5 K to room temperature, indicating Curie temperatures well above room temperature and inaccessible by our available instrumentation.

A comparison between the in-plane and out-of-plane magnetic hysteresis for the 0.2 ML Gd-doped 2DEG is shown in Fig. 4c. There are two surprising features in the magnetic anisotropy of these samples. First, a sharper (easier-axis) behavior is observed out-of-plane rather than in-plane indicating that the typical in-plane shape anisotropy for ferromagnetic thin films is not present. This observed crystalline anisotropy rules out the possibility of ferromagnetic dust contamination which is expected to be isotropic. The crystalline anisotropy is also in contrast to the previous reports from Ref. 8, where an easy-axis along $[1\bar{1}00]$ (perpendicular to the c-axis) was observed. It is, however, consistent with the possible role of threading dislocations. Secondly, even up to magnetic fields of 50 kOe, the in-plane magnetization does not reach the same saturation moment as observed in the out-of-plane orientation. This behavior is consistent with exchange coupled spins locked along the crystallographic c-axis that cannot be rotated in-plane. As shown in Tab. I and II, these samples exhibit a larger coercivity for in-plane orientation than out-of-plane.

The zero field cooled (ZFC) magnetization is plotted as a function of temperature in Fig. 4d for both Gd-doped 2DEGs. A non-vanishing spontaneous magnetization persists above room temperature. Moreover, a local peak in magnetization is observed in both samples at around 50 K. The temperature is remarkably close to the Curie temperature of cubic GdN (with a $T_C$ reported between 40 and 70 K depending on the growth conditions).[2,24,25] At the high Gd-delta-doping concentrations used, it is possible that a few GdN nanoprecipitates are present, but one would expect that to lead to a dip in magnetization just below $T_C$, not an excess as observed. Alternatively, oxygen contamination due to a small air leak into the sample space could also account for the local peak at 40 to 50 K due to the para-antiferromagnetic phase transition of solid oxygen.[26] However, this possibility is ruled out by examining the magnetization of the SiC wafer shown in Fig. 4d inset, where the raw magnetization data throughout the entire temperature range is around $1 \times 10^{-7}$ emu (approaching the detection limit of the instrument) with no significant peak at around 40 to 50 K. Here, the measurement on the SiC wafer is conducted with an out-of-plane magnetic field, and we employ the same measurement protocol for all samples. At the lowest temperatures, a clear paramagnetic contribution is observed as a steep rise in magnetization below 15 K consistent with a small concentration of paramagnetic spins. Therefore, we believe that the excess magnetization on observed in Fig. 4d is an intrinsic property of Gd-doped GaN, even if its physical origin is not elucidated.

Table I Magnetic properties with out-of-plane magnetic field

| Sample | T (K) | $M_s$ ($\mu_B$/Gd$^{3+}$) | $H_c$ (Oe) (scan direction) |
|---|---|---|---|
| $N_{Gd}^{2D}$ =0.2 ML | 5 | 41.1±0.87 | 72±134.6 (+→-) |
| | | | 113.4±107 (-→+) |
| | 300 | 29.8±1.6 | -74.4±47.8 (+→-) |
| | | | 45.2±211.7 (-→+) |
| $N_{Gd}^{2D}$ =0.4 ML | 5 | 9.2±1 | 64.3±255.4 (+→-) |
| | | | 203.2±275.1 (-→+) |
| | 300 | 7.6±1.6 | -142.6±277.2 (+→-) |
| | | | -161.8±244.4 (-→+) |

Table II Magnetic properties with in-plane magnetic field

| Sample | T (K) | $M_s$ ($\mu_B$/Gd$^{3+}$) | $H_c$ (Oe) (scan direction) |
|---|---|---|---|
| $N_{Gd}^{2D}$ =0.2 ML | 5 | 33.2±3 | -310.8±83.3 (+→-) |
| | | | 365.4±135.4 (-→+) |
| | 300 | 18.7±2.5 | -195.2±94.4 (+→-) |
| | | | 186.3±283.5 (-→+) |
| $N_{Gd}^{2D}$ =0.4 ML | 5 | 7.4±0.1 | -325.3±75.3 (+→-) |
| | | | 289.3±44.4 (-→+) |
| | 300 | 2.2±0.4 | -282.6±56.8 (+→-) |
| | | | 156.4±103.4 (-→+) |

B. Hall effect and magnetoresistance

Resistivity and Hall measurements are carried out on the ferromagnetic GaN 2DEGs using a Quantum Design PPMS 7T Model 6000. The Hall-bar geometry, based on ASTM F76 1-3-3-1 design with downscaled dimensions, is employed.[27] The measurement configuration is shown in the inset of Fig. 7a bottom right inset, where $I_{xx}$ is the injection current, $V_{xy}$ is the measured Hall voltage, and $V_{xx}$ is measured longitudinal voltage. The injection current $I_{xx}$ is set to be 10 $\mu$A for all the electrical measurements to minimize the self-heating in the 2DEG conduction channel. Prior to the Hall measurements, a leakage test is performed to ensure the absence of conduction through the regrowth interface. The current-voltage measurements taken before and after removal of the 2DEG are shown in Fig. 5. The leakage current is more than 5 orders of magnitude smaller than current flow through the 2DEG, therefore ensuring that charge transport occurs solely within the 2DEG channel.

An example of raw Hall resistivity $\rho_{xy}$ measured on the sample doped with 0.4 ML Gd at 20 K is shown in Fig. 7a top left inset. The $\rho_{xy}$ measured at high positive/negative field follows a linear behavior attributed to the ordinary Hall effect (OHE) in all samples. The temperature dependent carrier concentration and mobility calculated using the slope of the Hall resistivity and zero field resistivity are shown in Fig. 6. They are in good agreement with previous studies conducted on similar structures.[28-31] Non-linearity is observed in the $\rho_{xy}$ at low magnetic fields evidenced by the offset between the two linear regression fitting curves. The non-linear component of the Hall resistivity $\rho_{xy}^{NL}$, is obtained by subtracting the OHE component from the $\rho_{xy}$ and is plotted in Fig. 7a main panel for all the samples. $\rho_{xy}^{NL}$ from the control samples shows a similar field dependence as the two ferromagnetic Gd-doped 2DEGs. Further, the $\rho_{xy}^{NL}$ reaches half of the saturation value at fields between 5 kOe and 10 kOe at 20 K for all the compared samples, with or without Gd-doping or ferromagnetism. It is evident that the defect-induced ferromagnetic spins present in the Gd-doped 2DEGs do not significantly couple to the electrons in the 2DEG despite their spatial proximity. This finding is more clearly illustrated by plotting the magnetization $m$ and $\rho_{xy}^{NL}$ data on the same plot (Fig. 7d) at the same temperature 20 K. In conducting ferromagnets, and in ferromagnetic semiconductors, $\rho_{xy}^{NL}$ obeys,[32]

$$\rho_{xy}^{NL} = \rho_{xy}^{AHE} = c \times \rho_{xx}^{\gamma} \times m \qquad (2)$$

where $c$ is a temperature-independent proportionality constant, $\gamma$ is a power constant that could either be 1 (skew-scattering) or 2 (side-jump) depending on the origin of the scattering and $m$ is the out-of-plane magnetization of the sample. For the 0.2 ML Gd-doped 2DEG (Fig. 7d), the non-linear part ($\rho_{xy}^{NL}/\rho_{xx}^{\gamma}$ with $\gamma$=1 or 2) do not fit the magnetization data, which exhibit saturation at lower field. In contrast, the $\rho_{xy}^{NL}/\rho_{xx}^{\gamma}$ does not saturate until 50 kOe. The lack of a ferromagnetic-like response in the non-linear component indicates that in the ferromagnetic 2DEGs, the defect-induced ferromagnetic phase must be spatially separated from the 2DEG despite their intentional direct spatial overlap, thus disabling efficient magnetic coupling. This could occur only if the 2DEG electrons were laterally depleted in the regions where ferromagnetic spins were present. Since the 2D electron density is on the order of $10^{13}$ cm$^{-2}$, acceptor-like point defects would be unable to fully deplete the 2DEG. Therefore, we hypothesize that the ferromagnetic spins are present in larger scale defect clusters, such as threading dislocations ($\sim 10^9$ cm$^{-2}$ in our samples, see Sec. II B), that are known to behave as deep acceptors with significant electron depletion widths.[33,34] A previous study on non-magnetically doped AlGaAs/GaAs two-dimensional hole gas reported a similar non-linear Hall resistance and concluded that it is a genuine anomalous Hall effect that arose from skew scattering of spin-polarized charge carriers induced by Zeeman splitting between the spin-up and spin-down conduction channels.[35] This effect cannot explain our data, since, in our control samples, the non-linear Hall slope is present at magnetic fields where the Zeeman splitting is insignificant compared with $k_B T$ (Zeeman splitting for a g-factor of 2 is 1.3 K/T). An alternative explanation for the shape of $\rho_{xy}$ is provided in Sec. IV.

A comparison of the 5 K magnetization and 20 K $\rho_{xy}^{NL}/\rho_{xx}^{\gamma}$ data for the control samples and 0.4 ML Gd-doped samples is shown in Fig. 7b, c and e. Because the Hall resistivity of all the samples resembles a Brillouin function, one might suppose that it arises due to anomalous Hall effect within paramagnetic 2DEGs. However, this possibility is ruled out on two grounds. First, given the location of the Gd-doping within the 2DEGs and the induced anisotropic ferromagnetic phase, the lack of a ferromagnetic response in the Hall voltage (and the apparent paramagnetic-like response) indicates a lack of anomalous Hall effect. Secondly, the paramagnetic response from magnetization measurements follows the $1/T$ Curie law temperature dependence, which is distinct from that of the Hall resistivity, as shown in Fig. 8, indicating that the two responses (magnetization and Hall resistivity) are unrelated.

Magnetoresistance measurements are carried out on the same Hall bars. Raw sheet magnetoresistance $\rho_{xx} = \dfrac{V_{xx}}{I_{xx}}\dfrac{w}{l}$ (in the unit of $\Omega$/Square) of all samples is plotted as a function of magnetic field in Fig. 9a at 20 K. The general trend of the $\rho_{xx}$ of all the samples is a negative magnetoresistance with a fast decrease in the low field region and a linear decrease in the high field region suggesting that two different magnetic field dependent scattering mechanisms occur in all samples. Owing to the low mobility associated with the Gd-induced defect the 0.4 ML Gd-doped sample shows a zero-field sheet resistance ($\rho_{xx}$=1158 $\Omega$/Square) that is higher than both control samples ($\rho_{xx}$=412 and 488 $\Omega$/Square) as well as the 0.2 ML Gd-doped sample ($\rho_{xx}$=427 $\Omega$/Square). The slightly lower zero-field

$\rho_{xx}$ observed in the 0.2 ML Gd-doped sample, compared to the Control Sample 2, agrees with the fact that the 0.2 ML doped sample possesses a higher carrier concentration (see Sec. IV). The discrepancy in the zero-field sheet resistance between control samples is attributed to the difference in mobility (see Sec. IV). Fig. 9b shows the magnetoresistance $MR = \frac{\rho_{xx}(H) - \rho_{xx}(H=0)}{\rho_{xx}(H=0)}$ (in the unit of %) for all samples. Control Sample 1 and 2 show -4.5% and -3% changes in MR at a field value of 50 kOe, respectively. While -3.7% and -3.3% changes in MR are found for 0.2 ML and 0.4 ML Gd-doped samples, respectively. The MR of the control samples and the 0.2 ML Gd-doped 2DEG are quite similar. In contrast, the 0.4 ML Gd-doped 2DEG exhibits a fast decrease in resistance at low magnetic fields with a roll over at higher magnetic fields suggesting that this sample experiences a different combination of magnetic scattering processes. The MR of all the samples shows a linear decrease in the intermediate to high field region as shown in Fig. 9b. A detailed discussion of the magnetoresistance and its magnetoconductivity counterpart are presented in Sec. IV below.

### IV. Discussion

#### A. Nonlinear Hall slope

We now turn to the origin of the non-linearity in Hall slope observed in all samples. A suitable explanation for the non-linear Hall behavior in our data is the mixing of the magnetoconductivity $\sigma_{xx}$ and Hall resistivity $\rho_{xy}$ tensors. This can be significant in semiconductors, particularly in 2D transport, when there is a magnetoresistance, here due to the weak localization effect. The Hall conductivity can then become less import than the longitudinal magnetoconductivity. In the presence of a transverse magnetic field along the z-axis ($B_z = B \approx \mu_0 H$, $\mu_0$ is the vacuum permeability), the transverse (Hall) resistivity term $\rho_{xy}$ and the longitudinal electrical (magnetoresistance) resistivity term $\rho_{xx}$ are related to the off-diagonal and diagonal elements of the conductivity tensor, $\sigma_{xy}$ and $\sigma_{xx}$. The relationship between them is given by[36]

$$\rho_{xx} = \frac{\sigma_{xx}}{\sigma_{xx}^2 + \sigma_{xy}^2}, \quad \rho_{xy} = \frac{\sigma_{xy}}{\sigma_{xx}^2 + \sigma_{xy}^2} \quad (3)$$

where $\sigma_{xx} = \frac{ne\mu}{1+\mu^2 B^2}$ and $\sigma_{xy} = \frac{-ne\mu^2 B}{1+\mu^2 B^2}$ ($n$ is the electron concentration and $\mu$ is the mobility). Here, the measured physical quantities $\rho_{xx}$ and $\rho_{xy}$ equal to $\frac{1}{ne\mu}$ and $\frac{B}{ne}$, respectively. However, as shown in Fig. 9a, $\rho_{xx}$ exhibits negative magnetoresistance and multiple distinct regions, a magnetic field dependent correction term $\delta\sigma_{xx}$ should be included in $\sigma_{xx}$ where $\sigma_{xx} = \frac{ne\mu}{1+\mu^2 B^2} + \delta\sigma_{xx}$. In this case, $\rho_{xy}$ becomes strongly affected by the magnetoconductivity $\sigma_{xx}$ and is no longer proportional to the magnetic field. We calculate, in the appendix, a functional form for the term $\delta\sigma_{xx}$ that is based on the existence of weak localization, as indicated by the magnetoresistance in Fig. 9a. The conductivity tensors $\sigma_{xx}$ and $\sigma_{xy}$ are then transformed by matrix inversion to $\rho_{xx}$ and $\rho_{xy}$. A comparison between the calculated (solid lines) and measured (circles) $\rho_{xx}$, $\rho_{xy}$ and $\rho_{xy}^{NL}$ of Control Sample 2 at 20 K is shown in Fig. 10a, b and c (plotted in H field) as an example. The calculated $\rho_{xx}$, $\rho_{xy}$ and $\rho_{xy}^{NL}$ shows a relatively good agreement with the measured values. The carrier concentration $n$ and mobility $\mu$ for all the samples at 20 K is listed in Tab. III. These values are only slightly different than those calculated using the zero field resistivity, as in Fig. 6.

Table III Carrier concentration $n$ and mobility $\mu$ for all the samples at 20 K

| Sample | $n$ (/cm$^2$) | $\mu$ (cm$^2$/Vs) |
|---|---|---|
| Control Sample 1 | $3.3 \times 10^{13}$ | 564 |
| Control Sample 2 | $2.9 \times 10^{13}$ | 513 |
| Sample with $N_{Gd}^{2D}$ =0.2 ML | $4.4 \times 10^{13}$ | 407 |
| Sample with $N_{Gd}^{2D}$ =0.4 ML | $3.1 \times 10^{13}$ | 195 |

The carrier concentration and mobility of Gd-doped samples at 300 K are measured from the Hall slope and zero-field resistivity. No non-linearity of $\rho_{xy}$ is observed Fig. 11a. This is clearly contrast with what is observed at 20 K (Fig. 7a inset) and results from the negligible magnetoresistance at 300 K as shown in Fig. 11b. The carrier concentrations and mobility obtained are listed in Table IV. As expected for polarization-induced 2DEG, there is no significant change in carrier density from room temperature to low temperature. Additionally, all samples have carrier concentrations within 34% indicating that Gd-doping has no obvious effect on the carrier concentration, as expected for an isoelectronic dopant.

Table IV Carrier concentration $n$ and mobility $\mu$ for all samples at 300 K

| Sample | $n$ (/cm$^2$) | $\mu$ (cm$^2$/Vs) |
|---|---|---|
| Control Sample 1 | $2.4 \times 10^{13}$ | 454 |
| Control Sample 2 | $2.3 \times 10^{13}$ | 445 |
| Sample with $N_{Gd}^{2D}$ =0.2 ML | $3.1 \times 10^{13}$ | 335 |
| Sample with $N_{Gd}^{2D}$ =0.4 ML | $2.7 \times 10^{13}$ | 198 |

B. 2DEG electronics

The deduced carrier concentration of the control samples ($2.4 \times 10^{13}$ and $2.3 \times 10^{13}$ cm$^{-2}$) are lower than the electron concentration of $4.2 \times 10^{13}$ cm$^{-2}$ calculated from the 1D Poisson-Schrodinger solver in Sec. II at 300 K. This reduction of 2DEG concentration could be explained by the fact that the AlN layer is partially strain relaxed leading to a reduced piezoelectric polarization. Additionally, since the band-gap of AlN tends to shrink when subjected to biaxial tensile strain, then, the conduction band offset between the AlN and GaN of 2.1 eV used in the simulation may be overestimated. Assuming all the band gap shrinkage contributes to the reduction of the conduction band offset, the amount of reduction in $\Delta E_c^0$ is calculated as,[37]

$$\Delta E_c^0 - \Delta E_c = -[d_1 e_\perp + 2 d_2 e_{//}] \quad (4)$$

where $e_\perp$ and $e_{//}$ are the uniaxial and biaxial strain, $d_1$ and $d_2$ are the deformation energy and $\Delta E_c$ is the conduction band offset when AlN is stained.[38] The electron effective mass in GaN also tends to increase due to the nature of the non-parabolic energy dispersion in triangular well which follows the Ando formula,[39]

$$\frac{\Delta m^*}{m^*} \approx \sqrt{1 + 4 \times \frac{<K>_i + E_F}{E_g}} - 1 \quad (5)$$

where $E_F$ is the Fermi energy measured from the bottom of the lowest sub-band and $<K>_i$ is the kinetic energy of the motion perpendicular to the interface and can be calculated as $<K>_i = E_i / 3$ for the triangular well where $E_i$ is the subband energy measured from the conduction band energy minimum at the heterointerface. Here, $<K>$ is calculated using the ground subband energy $E_1$ since most of the electrons reside on the ground subband. A self-consistent 1D Poisson-Schrodinger solution on a fixed structure (5nm AlN/300nm GaN) including the following variables: relaxation of AlN layer ($R$, $R$=0 or $R$=1 meaning fully strained or fully relaxed), conduction band offset between the AlN and GaN ($\Delta E_c$) and electron effective mass ($m^*$) is obtained in order to match the carrier concentration. It is found that the combination of of $R$=68%, $\Delta E_c$=1.93 eV, $m^*$=0.24$m_0$ and $R$=74%, $\Delta E_c$=1.96 eV, $m^*$=0.24$m_0$ ($m_0$ is the electron rest mass) matches best with the calculated carrier concentration of Control Sample 1 and 2, respectively. The fact that Control Sample 1 possesses a slightly less relaxed AlN capping layer agrees with its slightly higher carrier concentration and mobility as shown in Tab. III compared to Control Sample 2.

Doping with Gd induces more carriers in the 2DEG in 0.2 ML Gd-doped sample, while the 0.4 ML Gd-doped sample exhibits an electron concentration relatively similar to the control samples at 20 K and 300 K. The relatively high electron concentration in 0.2 ML Gd-doped sample does not contradict the presence of acceptor-like defects due to Gd-doping as reported by Bedoya-Pinto et al.,[23] since the presence of Gd dopants in the vicinity of the heterointerface could alter the local crystal potential field distribution and conduction band offset $\Delta E_c$ at the heterointerface leading to a heavier electron effective mass and deeper quantum well. This could substantially increase the 2D carrier density and overwhelm the effect from the presence of Gd associated acceptor-like defects within the quantum well. However the quantitative change in band alignment and electron effective mass and how they depend on the Gd-doping concentration needs to be further studied. For these reasons, the band diagrams of the Gd-doped samples are not attempted here.

However, the observation of partial strain-relaxation in control samples suggests a similar situation existing in Gd-doped samples. The simulated band diagram and carrier concentration profile adopting the new physical parameters of Control Sample 2 are shown as an example and depicted as the solid line in Fig. 2b. The reduced band bending in the AlN layer and the shallower quantum well directly resulting from the reduced piezoelectric charge density and conduction band offset gives a lower 2DEG concentration. The mobility exhibits a reduction as the Gd-doping concentration increases, which is consistent with enhanced Coulomb scattering due to Gd-doping generated defects.

## V.     Conclusions

Polarization-induced AlN/GaN 2DEGs doped with Gd exhibit a defect-induced ferromagnetism with a Curie temperature above room temperature and an anisotropic magnetization. Sharper switching is observed with the field aligned along the c-axis rather than within the basal plane. Surprisingly, the saturation magnetization for in-plane orientation is lower than for out-of-plane orientation indicating a strong pinning of the defect-spins associated with the magnetism along the polar direction of GaN. The saturation magnetization is larger than what is possible from $Gd^{3+}$ spins alone indicating the presence of a large concentration of spin-polarized defects that generate the ferromagnetic response. In non-Gd-doped samples, weak ferromagnetism and paramagnetism are also observed possibly caused by the difference in the density of spin polarized defects. Hall measurements of ferromagnetic Gd-doped 2DEGs exhibit non-linearity in the Hall resistivity that does not track the magnetization. The non-linear Hall effect is consistent with the mixing of the conductivity tensor $\sigma_{xx}$ into the Hall resistivity $\rho_{xy}$. The lack of coupling between the ferromagnetic phase intentionally placed within the 2DEG indicates that the electrons in the 2DEG must be depleted from the ferromagnetic region. We hypothesize that the depletion is induced by high densities or clusters of acceptor-like defect spins. The strain-relaxation is associated with the formation of threading dislocations oriented along the c-axis, which are known to cause both electron depletion and charge scattering. Therefore such dislocation cores could account for the anomalous ferromagnetic phase and its lack of coupling to a 2DEG. It remains to be seen if these acceptor-like defect spins might couple to the valence band of GaN, studies which necessitate the same measurements need to be carried out in p-GaN or 2D hole gasses.

**Acknowledgements**

The authors thank the staff C. Selcu at NanoSystems Laboratory (NSL) and D. Ditmer at NanoTech West (NTW) at the Ohio State University for technical assistance. Dr. Digbijoy N. Nath and Dr. Siddharth Rajan provided the 2DEG growth recipe. This work was supported by the National Science Foundation under Grant No. CBET-1133589.

**Appendix**

Here, we calculate a functional form for the term $\delta\sigma_{xx}$. In the low field region (< 5 kOe), the fast decrease in magnetoresistance is induced by the weak-localization effect, as suggested by previous reports, which leads to a quantum correction to $\sigma_{xx}$ given by[40]

$$\delta\sigma_{xx}^{WL} = \sigma_0[\psi(\frac{1}{2}+\frac{\hbar}{4eDB\tau_i}) - \psi(\frac{1}{2}+\frac{\hbar}{4eDB\tau_e}) + \ln(\frac{\tau_i}{\tau_e})] \quad (6)$$

where $\sigma_0 = e^2/2\pi^2\hbar$, $\psi$ is the digamma function, $D$ is the electronic diffusion constant given by $D = v_F^2\tau_e/2$ where $v_F$ is the Fermi velocity, and $\tau_e$ and $\tau_i$ are the elastic and inelastic scattering rates, respectively. With the new conductivity tensor, $\sigma_{xx} = \frac{ne\mu}{1+\mu^2 B^2} + \delta\sigma_{xx}^{WL}$, we find that the calculated $\rho_{xx}$ fits the data well in the low field region, but shows saturation at high field, which does not agree with the observed linear decrease (Fig. 9b). The deviation suggests that an additional correction term to the high field $\sigma_{xx}$ has to be included. Previous reports suggest that, in 2D transport, electron-electron interaction would lead to a negative correction to $\sigma_{xx}$ and a continuous decrease in $\rho_{xx}$.[40,41] The theory predicts that the amount of reduction $\delta\sigma_{xx}^{EEI}$ is a field independent term when $k_B T \tau_e < 1$. The detailed expression of $\delta\sigma_{xx}^{EEI}$ is not evaluated here since there is a possibility of introducing a similar constant offset in $\sigma_{xx}$ from the misalignment in the measurement geometry (i.e. Hall bar detection leg has a finite width and is not an ideal point contact). We use a single variable $\delta\sigma_{xx}^{EEI}$ to account for the combined effect from the electron-electron interaction and the possible misalignment. The conductivity tensor now becomes, $\sigma_{xx} = \frac{ne\mu}{1+\mu^2 B^2} + \delta\sigma_{xx}^{WL} + \delta\sigma_{xx}^{EEI}$, and after the matrix inversion, $\rho_{xx}$ experiences a $B^2$ dependent parabolic decrease in the high field region. This still does not yield a linearly decreasing $\rho_{xx}$ (Fig. 9b). For a better fitting between the calculated and measured $\rho_{xx}$ in the high field region, we attempt to introduce a second weak localization correction $\delta\sigma_{xx}^{WL2}$ which could be induced by a different type of defect. The elastic $\tau_{e2}$ is much faster for $\delta\sigma_{xx}^{WL2}$ than for $\delta\sigma_{xx}^{WL}$, altering the curvature of $\rho_{xx}$ in the high field region from a parabolic to a linear decrease. $\sigma_{xx}$ and $\sigma_{xy}$ are now given by

$$\sigma_{xx} = \frac{ne\mu}{1+\mu^2 B^2} + \delta\sigma_{xx}^{WL} + \delta\sigma_{xx}^{EEI} + \delta\sigma_{xx}^{WL2}, \quad \sigma_{xy} = \frac{-ne\mu^2 B}{1+\mu^2 B^2} \quad (7)$$

There is no correction to $\sigma_{xy}$ from either weak localization or electron-electron interaction. Eqn. 7 is used to calculate $\rho_{xx}$ and $\rho_{xy}$ which are then fitted to the measured values using least-squares fitting method. The variables $n$, $\mu$, $\delta\sigma_{xx}^{EEI}$, $4eDB\tau_e$, $4eDB\tau_i$, $4eDB\tau_{e2}$ and $4eDB\tau_{i2}$ can be deduced from the fitting process described above where $n$ and $\mu$ are listed in Tab. III. Since we focus on demonstrating the mixing of $\sigma_{xx}$ and $\rho_{xy}$ as the cause of the non-linearity in the Hall signal, the discussion of $\delta\sigma_{xx}^{EEI}$, $4eDB\tau_e$, $4eDB\tau_i$, $4eDB\tau_{e2}$ and $4eDB\tau_{i2}$ is beyond the scope of this paper, and thus is not discussed here.

**Figure Captions**

Figure 1 (Color online) (a) SIMS characterization of the five major Gd isotopes concentration (thin colored line) and average concentration (thick black line) of the Gd:GaN calibration stack grown with different Gd effusion cell temperatures. (b) Average Gd-doping concentration of the Gd:GaN calibration stack as a function of the reciprocal of the Gd cell temperature and the fitted curve using $A \times e^{-b/T}$ function. (c) Derived Gd-doping rate as a function of Gd cell temperature and the fitted curve using $A \times e^{-b/T}$ function.

Figure 2 (Color online) (a) Schematic diagram of the Gd-delta-doped samples. (b) Simulated band diagram and carrier concentration profile of the Control Sample 2 using different physical parameters. (c) SIMS characterization of the 0.2 ML Gd-doped sample.

Figure 3 (Color online) (a) High resolution X-ray diffraction $\omega - 2\theta$ scan of the Control Sample 1. Inset: $\omega$-rocking curves of the (002) and (102) diffraction peaks. AFM of the (b) 0.2 ML and (c) 0.4 ML Gd-doped sample.

Figure 4 (Color online) (a) Background corrected magnetization hysteresis loops of the SiC wafer, control samples and Gd-doped samples with out-of-plane magnetic field at 300 K. Inset: detailed comparison of magnetization between the control samples and SiC wafer at 300 K (top left) and 5 K (bottom right). (b) Background corrected magnetization hysteresis loops of the 0.2 ML and 0.4 ML Gd-doped samples at 5 K and 300 K. (c) Background corrected magnetization hysteresis loops of the 0.2 ML Gd-doped sample with out-of-plane

(parallel to [0001]) and in-plane (perpendicular to [0001]) magnetic field. (d) Background corrected magnetization of the Gd-doped samples after zero-field cooling with in-plane magnetic field. Inset: raw magnetization of the 0.4 ML Gd-doped sample and SiC wafer.

Figure 5 (Color online) Comparison of the IV characteristic in log scale (main panel) and linear scale (right inset) before and after removing of the 2DEG conduction channel of the Control Sample 1. Left inset: the sample structures used in the leakage test.

Figure 6 (Color online) Temperature dependent (a) carrier concentration and (b) mobility of the control samples and Gd-doped samples calculated from the Hall measurement (without taking MR mixing effect).

Figure 7 (Color online) (a) Comparison of $\rho_{xy}^{NL}$ of the control samples and Gd-doped samples. Top right inset: raw $\rho_{xy}$ (blue dot) with high field linear fit (pink dash) of the 0.4 ML Gd-doped sample at 20 K. Bottom left inset: Hall-bar geometry employed in the Hall measurement. Comparison between $-\rho_{xy}^{NL}/\rho_{xx}^{\gamma}$ at 20 K and the diamagnetic background corrected magnetization at 5 K of the (b) Control Sample 1 (c) Control Sample 2 and (e) 0.4 ML Gd-doped sample. Comparison between $-\rho_{xy}^{NL}/\rho_{xx}^{\gamma}$ and the diamagnetic background corrected magnetization at the same temperature 20 K of the (d) 0.2 ML Gd-doped sample. $\gamma$ is the power constant that could either be 1 or 2 depending on the origin of the scattering.

Figure 8 (Color online) Comparison between the temperature dependent $\Delta\rho_{xy}^{NL}$ and Brillouin function ($S = \frac{3}{2}$, $g = 4.5$) of the control samples and Gd-doped samples.

Figure 9 (Color online) Comparison of (a) $\rho_{xx}$ ($\Omega$/Square) and (b) MR (%) with linear fit in the high field region of the control samples and Gd-doped samples at 20 K.

Figure 10 (Color online) Comparison between the calculated (red dash line) and measured (black open circles) (a) $\rho_{xy}$ (b) $\rho_{xx}$ and (c) $\rho_{xy}^{NL}$ of the Control Sample 2 at 20 K.

Figure 11 (Color online) (a) Hall resistivity $\rho_{xy}$ (wine cross) of the 0.2 ML Gd-doped sample at 300 K with linear regression fit of $\rho_{xy}$ in the high positive (red dotted line) and negative (blue dotted line) magnetic field region. (b) Comparison of the magnetoresistance $\rho_{xx}$ of the 0.2 ML Gd-doped sample at 20 K and 300 K.

**Figure 1 (One column)**

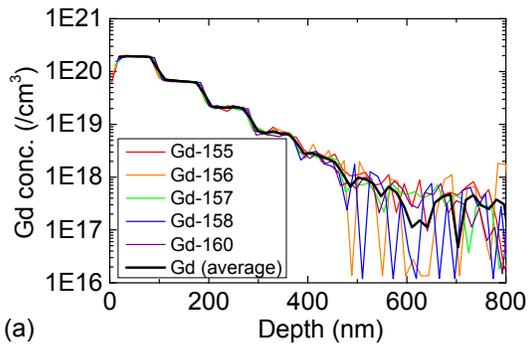

(a)

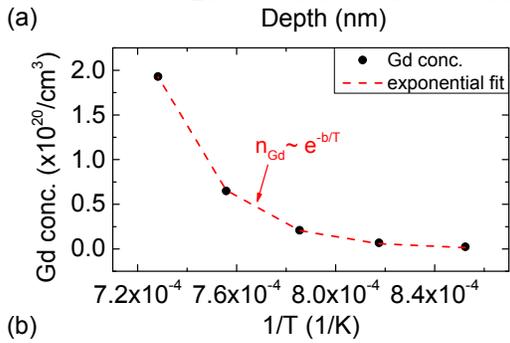

(b)

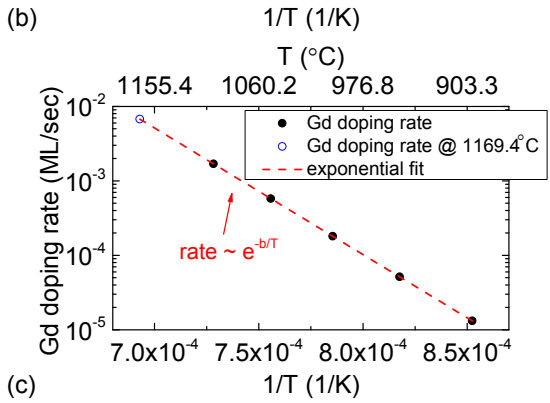

(c)

**Figure 2 (One column)**

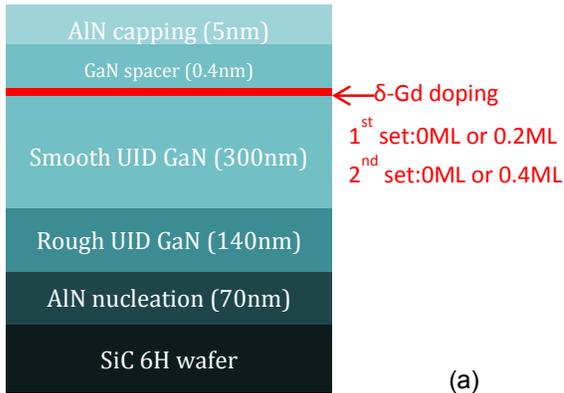

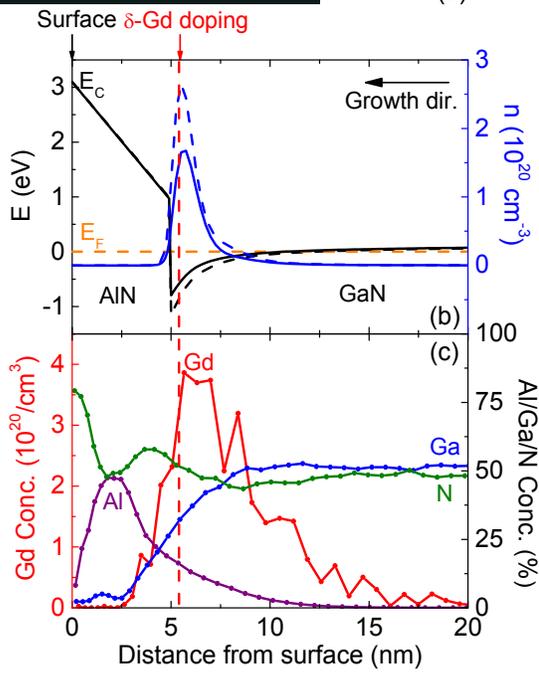

**Figure 3 (One column)**

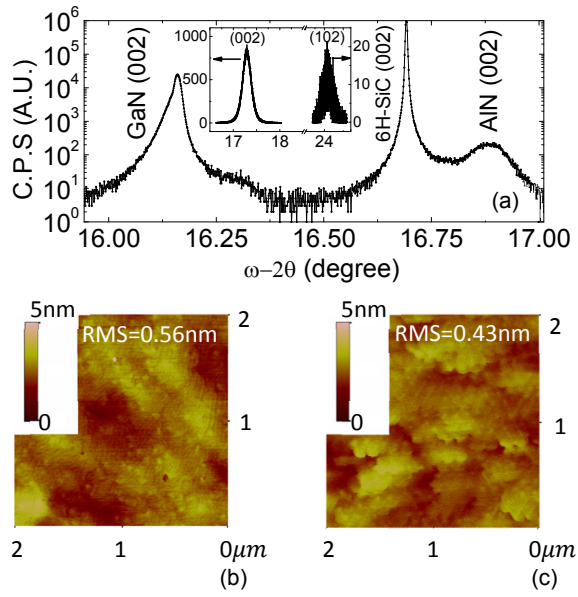

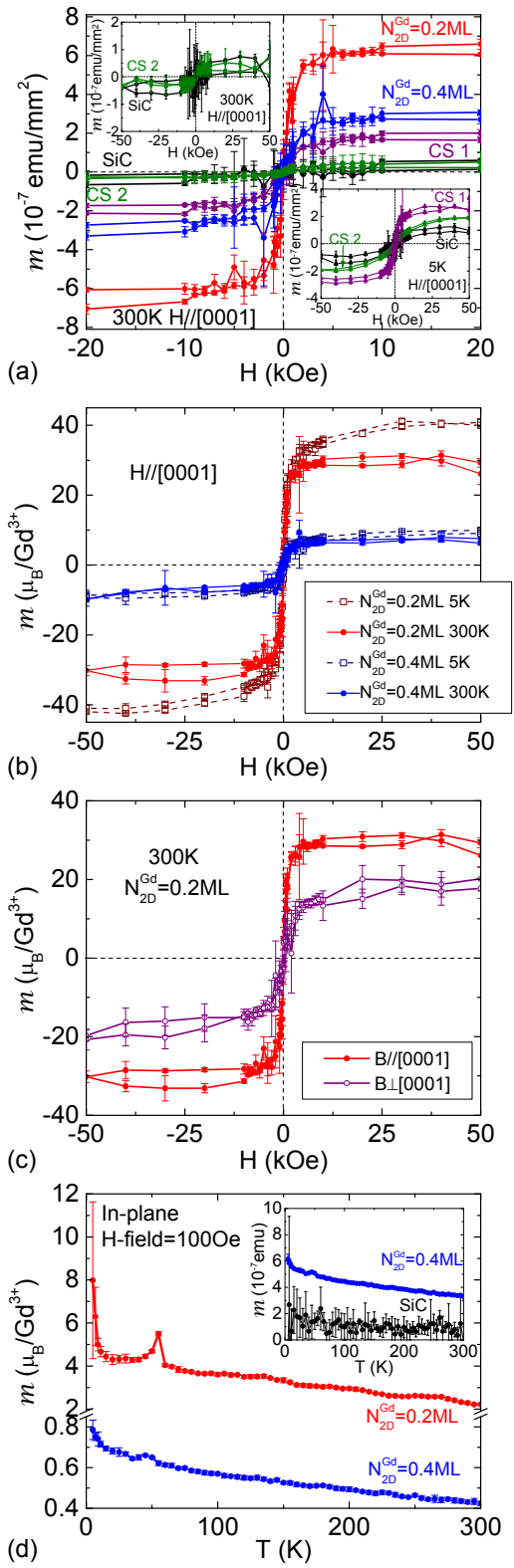

Figure 5 (One column)

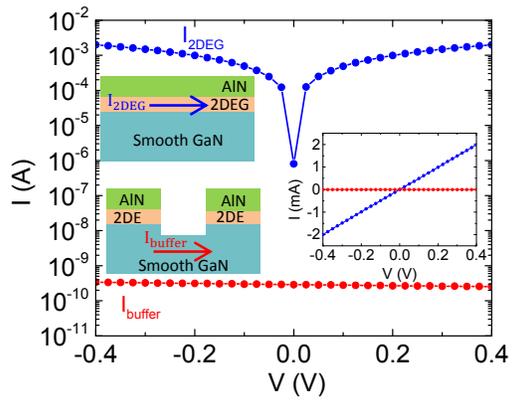

**Figure 6 (One column)**

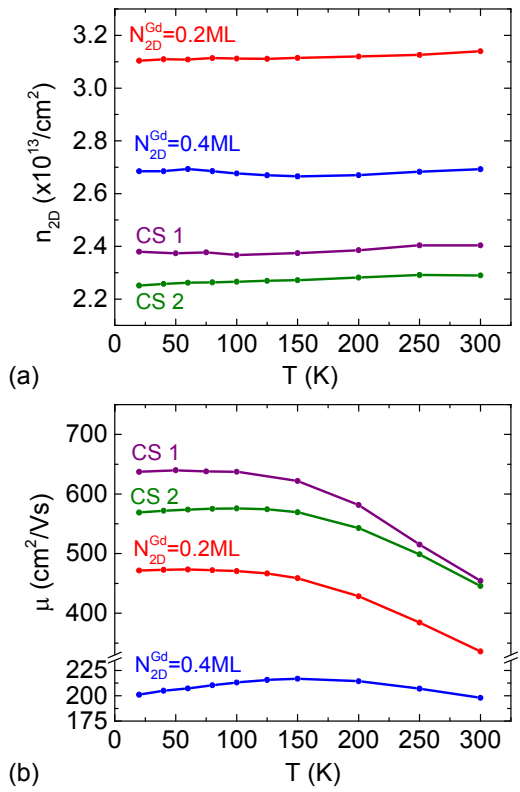

**Figure 7 (One column)**

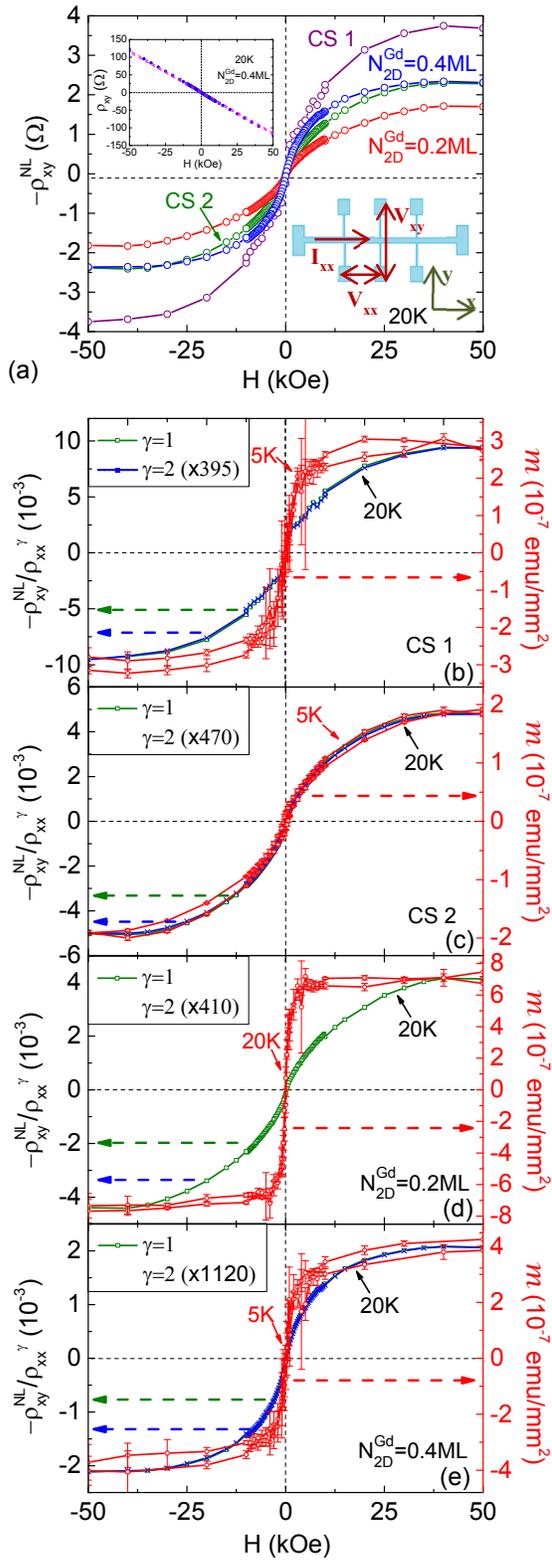

**Figure 8 (One column)**

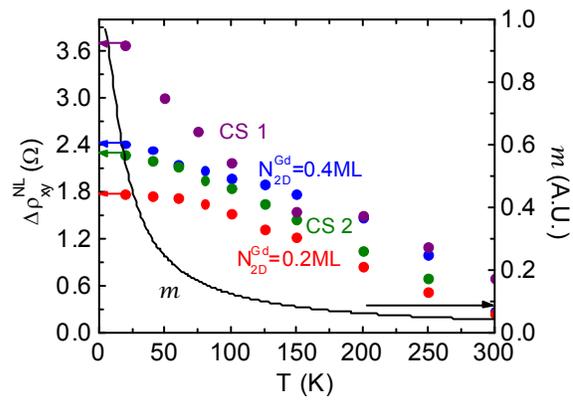

**Figure 9 (One column)**

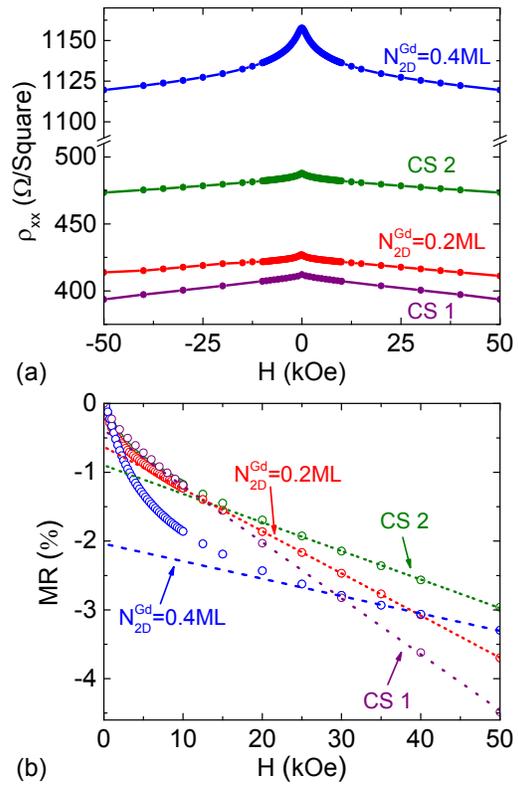

(a)

(b)

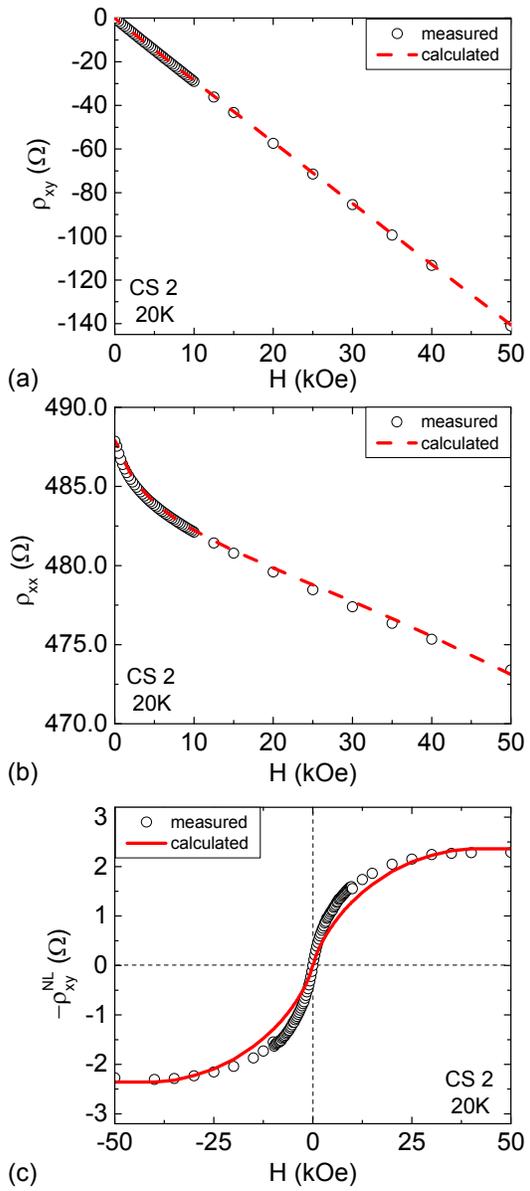

(a)

(b)

(c)

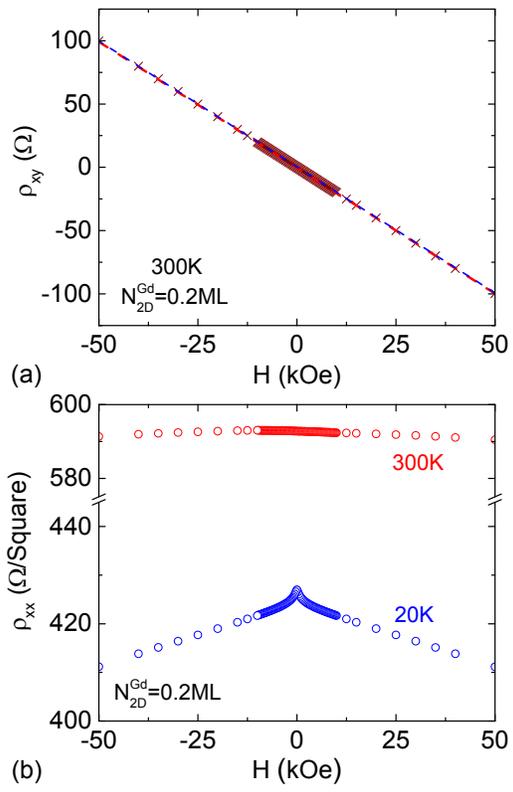

(a)

(b)